\documentclass[doublecol]{epl2} 

\usepackage{amsfonts}
\usepackage{amsmath}
\usepackage{amssymb}
\usepackage{hyperref}
\hypersetup{colorlinks=true,linkcolor=blue,citecolor=blue,filecolor=blue,urlcolor=blue,breaklinks=true}

\title{Effects of the Lorentz symmetry violation on relativistic neutral scalar bosons: Scattering and bound states}
\shorttitle{Effects of the LSV on relativistic scalar particles} 
\author{Luis B. Castro\inst{1} \and Antonio S. de Castro\inst{2}}
\shortauthor{L.B. Castro \and A.S. de Castro }
\institute{
\inst{1} Departamento de F\'{\i}sica, Universidade Federal do Maranh\~{a}o (UFMA),
  Campus Universit\'{a}rio do Bacanga,
  65080-805, S\~{a}o Lu\'{\i}s, MA, Brazil\\ \email{lrb.castro@ufma.br}\\ 
\inst{2} Departamento de F\'{\i}sica,
Universidade Estadual Paulista (UNESP), Campus de Guaratinguet\'{a}, 12516-410 Guaratinguet\'{a}, S\~ao
Paulo, Brazil\\
\email{antonio.castro@unesp.br}
}

\abstract{In this letter, we investigate the effects of Lorentz symmetry violation on a relativistic neutral scalar boson within the framework of the Klein-Gordon formalism. We consider a tensor $(K_F)_{\mu \nu \alpha \beta}$ out of the Standard Model Extension, which describes the field configuration consisting of a constant magnetic field $\vec{B}=B\hat{z}$ and a cylindrical electric field $\vec{E}=\frac{\lambda}{r}\hat{r}$. We analyze and discuss the effects of Lorentz symmetry violation on the equation of motion and show that the presence of a specific Lorentz symmetry violation parameter is essential to obtain analytical solutions for scattering and bound states. Employing the partial wave approach in cylindrical coordinates, we calculate the relativistic phase shift, scattering amplitude and $S$-matrix, and discuss the effects of Lorentz symmetry violation on the phase shift. Furthermore, we obtain bound-state solutions by examining the poles of the $S$-matrix and compare our findings with those reported in the literature. Our results reveal that bound-state solutions are only feasible for a restrict range of Lorentz symmetry violation parameters, which contradict previous studies.}

\pacs{11.30.Cp}{Lorentz and Poincaré invariance}
\pacs{11.80.-m}{Relativistic scattering theory}
\pacs{03.65.Ge}{Solutions of wave equations: bound states}

\begin{document}

\maketitle

\section{Introduction}
\label{intro}

The study of Lorentz symmetry, a cornerstone of modern physics, plays a fundamental role in the Standard Model of particle physics and general relativity. However, exploring possible violations of this symmetry has garnered significant attention in recent years, particularly within the framework of the Standard Model Extension (SME) \cite{PRL63:224:1989,PRD39:683:1989,PRD40:1886:1989,PRL66:1811:1991,NPB359:545:1991,PRD51:3923:1995,PLB381:89:1996,PRD55:6760:1997,PRD58:116002:1998,PRD59:116008:1999}. The SME provides a theoretical platform to investigate subtle deviations from Lorentz symmetry, offering insights into potential physics beyond the Standard Model. Among its many facets, the SME introduces a variety of tensor fields that describe Lorentz-violating effects, enabling the study of their impact on different physical systems. The electromagnetic sector of the SME includes two Lorentz-violating contributions that affect the propagation of electromagnetic waves in spacetime: the CPT-odd sector \cite{PRD55:6760:1997,PRD58:116002:1998} and the CPT-even sector \cite{PRD41:1231:1990,PRL87:251304:2001,PRD66:056005:2002,PRL97:140401:2006}.

In this context, scalar fields have emerged as a valuable testbed for probing Lorentz symmetry violation (LSV) due to their mathematical simplicity and physical relevance in several domains, including quantum field theory, cosmology, and condensed matter systems. By analyzing the behavior of scalar particles under modified conditions, such as non-trivial electromagnetic fields or Lorentz-violating backgrounds, one can gain a deeper understanding of how such violations manifest and influence physical observables. The influence of the LSV on the dynamics of scalar particles via Klein-Gordon (KG) formalism has been widely discussed in the literature \cite{AP360:596:2015,AP373:115:2016,EPJP132:25:2017,EPJC78:999:2018,AP399:117:2018,ADHEP2019:8462973:2019,ADHEP2019:1248393:2019,IJMPA35:2050023:2020,EPJP135:123:2020,EPL136:41002:2021,EPL136:61001:2021,EPL139:30001:2022,EPL141:60002:2023}. However, all these studies focus on bound states, neglecting the important investigation of scattering states. Therefore, we believe this issue warrants further exploration.    

In this work, we focus on the relativistic KG formalism to study the dynamics of a neutral scalar boson in the presence of a Lorentz symmetry-violating background. Specifically, we consider the contribution of the tensor $(K_F)_{\mu \nu \alpha \beta}$, an essential component of the SME, to describe a field configuration comprising a constant magnetic field $\vec{B}=B\hat{z}$ and a cylindrical electric field $\vec{E}=\frac{\lambda}{r}\hat{r}$. This setup not only enables us to examine the interplay between scalar fields and electromagnetic fields but also provides a unique opportunity to investigate how Lorentz symmetry-violating parameters influence both scattering and bound-state phenomena. By employing the partial wave method in cylindrical coordinates, we analyze the relativistic phase shift, scattering amplitude, and
$S$-matrix, highlighting the effects of Lorentz symmetry violation on these quantities. Furthermore, we determine bound-state solutions by examining the poles of the $S$-matrix and compare our findings with existing results in the literature. Interestingly, we find that bound-state solutions are only achievable within a restricted range of Lorentz symmetry-violating parameters, contradicting previous studies and offering new insights into the role of these parameters in relativistic quantum systems.

This paper is structured as follows: the second Section provides an overview of the theoretical framework and presents the modified KG equation that incorporates LSV. In the third Section, we perform a detailed qualitative analysis of the effective potential, exploring the connections between the parameters of our problem and the regions where bound and scattering states are expected. The fourth Section focuses on deriving the scattering solutions, calculating the phase shift, scattering amplitude, and determining the $S$-matrix. In the fifth Section, we analyze the bound-state solutions, comparing our results with those found in the literature. Finally, the sixth Section offers a summary of our findings and discusses their potential implications for future research.

\section{Klein–Gordon equation  in the background of Lorentz symmetry violation}
\label{kg_equation}
 
We investigate the relativistic dynamics of a massive neutral scalar particle under the effects of LSV through a nonminimal coupling in the KG equation given by 
\begin{equation}
\hat{p}_\mu \hat{p}^\mu \rightarrow \hat{p}_\mu \hat{p}^\mu -\frac{g}{4}(K_F)_{\mu \nu \alpha \beta} F^{\mu \nu} F^{\alpha \beta} ,
\end{equation}
\noindent where $g>0$ is a coupling constant, $F^{\mu \nu}$ is the electromagnetic
tensor, and $(K_F)_{\mu \nu \alpha \beta}$ is a tensor that governs the LSV in CPT-even electrodynamics within the SME framework \cite{PRD81:105015:2010,PRD84:045008:2011}. In this context, the term CPT-even refers to coefficients that preserve CPT symmetry, meaning they remain invariant under charge conjugation ($C$), parity ($P$), and time reversal ($T$). Therefore, while indicating Lorentz Violation (LV), they do not violate CPT symmetry itself. This tensor has $19$ components, exhibiting the same symmetries as the Riemann tensor. These symmetries can be expressed in terms of four $3\times 3$ matrices, given by
\begin{equation}
    (\kappa_{DE})_{jk}=-2(K_F)_{0j0k},
\end{equation} 
\begin{equation}
    (\kappa_{HB})_{jk}=\frac{1}{2}(K_F)^{pqlm}\varepsilon_{jpq} \varepsilon_{klm} , 
\end{equation}
\begin{equation}
(\kappa_{DB})_{jk}=-(\kappa_{HE})_{kj}= (K_F)^{0jpq}\varepsilon_{kpq},
\end{equation}
\noindent where the even-parity sector is composed by the $\kappa_{D E}$ and $\kappa_{H B}$ symmetric matrices with $11$ independent components, and the odd-parity sector is composed by matrices without inherent symmetry $\kappa_{D B}$ and $\kappa_{H E}$ with $8$ components.   

In this way, the modified KG equation for neutral scalar bosons of mass $m$ under the effects of LSV is given by $(\hbar=c=1)$
\begin{equation}\label{eq2}
\begin{split}
&\hat{p}_\mu \hat{p}^\mu \psi +\frac{g}{2}(\kappa_{DE})_{jp} E^{j} E^{p} \psi+g(\kappa_{DB})_{j\omega} E_{j}  B^\omega \psi\\& -\frac{g}{2}(\kappa_{HB})_{bc} B^bB^c \psi = m^2 \psi\,,
\end{split}  
\end{equation}
\noindent where $E_{i}=F_{0i}$ and $B_{i}=\frac{1}{2}\epsilon_{ijk}F^{jk}$ represent the electric and magnetic fields, respectively. In this stage, it is useful to consider the current. The conservation law for $J^{\mu}$ follows from the standard procedure and it results in $\partial_{\mu}J^{\mu}=0$, where
\begin{equation}\label{jmu}
J^{\mu}=-\frac{1}{m}\mathrm{Im}\left( \psi^{\ast}\partial^{\mu}\psi \right)\,.
\end{equation}
\noindent The time component of $J^{\mu}$ is not positive definite, but $J^{0}$ may be interpreted as a charge density. This implies that the normalization condition can be expressed as $\int d\tau J^{0}=\pm 1$, where the plus (minus) sign must be used for a positive (negative) charge. From (\ref{jmu}), it can be seen that in this case the normalization condition is not affected by the presence of LSV, in contrast to another model of LSV \cite{EPL141:60002:2023}.   

Now, the background of Minkowski spacetime is considered in cylindrical coordinates
\begin{equation}\label{cilin}
ds^2=-dt^2+dr^2+r^2 d\varphi^2+dz^2\,,
\end{equation}
\noindent and in this case, it is possible to define the following physical scenario in which the LSV leads to a configuration for the crossed electric and magnetic fields \cite{JPG39:055004:2012,AP360:596:2015,EPJP132:25:2017,EPJP135:247:2020,EPL136:41002:2021}
\begin{equation}\label{campos}
\vec{B}=B\hat{z},\quad     \vec{E}= \frac{\lambda}{r} \, \hat{r} ,  
\end{equation}
\noindent where $B$ is a constant magnetic field, $\lambda$ is a linear electric charge density, and $\hat{z}$ and $\hat{r}$ are unit vectors in the $z$ and radial directions, respectively. Additionally, let us consider the following non-null components
of the tensor  $(K_F)_{\mu \nu \alpha \beta} $ given by:
\begin{eqnarray}
(\kappa_{DE})_{rr} &=& \kappa_1\,,\nonumber \\
(\kappa_{HB})_{zz} &=&\kappa_2\,,\label{kappa}\\
(\kappa_{DB})_{rz} &=&\kappa_3\,,\nonumber
\end{eqnarray}
\noindent where, $\kappa_1$, $\kappa_2$ and $\kappa_3$ are constants. The parameters $\kappa_{i}$ were introduced in cylindrical coordinates, rather than Cartesian coordinates. This choice was made because, in Cartesian coordinates, these parameters would no longer remain constant when transformed into cylindrical coordinates. Using (\ref{campos}) and (\ref{kappa}), the equation (\ref{eq2}) is rewritten in cylindrical coordinates as follows 
\begin{equation}
\begin{split}
    &\Big(-\frac{\partial^2}{\partial t^2}+\nabla^2 -\frac{\kappa_1 g}{2} \frac{\lambda^2}{r^2}-\kappa_3 g\frac{\lambda B}{r}+\frac{\kappa_2 g B^{2}}{2}-m^2\Big)\psi =0\,,
\end{split} \label{eq3}
\end{equation}
\noindent where $\nabla^2$ is the Laplacian operator. At this stage, we can adopt the usual decomposition 
\begin{equation}
 \psi(r)= e^{-i\varepsilon t } e^{i l \varphi} e^{i k_z z} \, \frac{\phi(r)}{\sqrt{r}}  , \label{ansatz}
\end{equation}
\noindent where $\varepsilon$ is the energy of the system, $l=0,\pm 1, \pm 2, \ldots$ are the eigenvalues of the angular momentum operator $L_{z}=-i\partial_{\varphi}$ and $k_z \in \left<-\infty,+\infty \right>$
is the quantum number associated with linear momentum in the $z$-direction. Substituting the solution (\ref{ansatz}) into equation (\ref{eq3}), we obtain the following radial equation
\begin{equation}\label{eq4}
    -\frac{d^2\phi(r)}{dr^2}+V_{\mathrm{eff}}\phi(r)=\Lambda^{2}\phi(r)\,,
\end{equation}
\noindent where
\begin{equation}\label{veff}
V_{\mathrm{eff}}=\frac{\delta}{r}+\frac{\gamma^2_l-\frac{1}{4}}{r^2}\,,
\end{equation}
\noindent with
\begin{equation}
    \Lambda^2=\varepsilon^2-m^2-k^2_z+\frac{\kappa_{2}gB^{2}}{2}\,,
\end{equation}
\begin{equation}\label{delta}
\delta=\kappa_{3}g\lambda B\,,
\end{equation}
\begin{equation}\label{gammal}
    \gamma_l=  \sqrt{\frac{\kappa_{1}g\lambda^2}{2}+l^2}\,.
\end{equation}
\noindent Meanwhile,
\begin{equation}\label{j0}
J^{0}=\frac{\varepsilon}{m}\frac{\vert\phi\vert^{2}}{r}\,.
\end{equation}
\noindent Therefore, the equation (\ref{eq4}) describes the relativistic dynamics of a massive neutral scalar particle under the effects of LSV. With $\phi(0)=0$ and $\int_{0}^{\infty}\vert \phi\vert^{2}dr<\infty$, this equation is precisely the time-independent radial Schrödinger equation for a Coulomb-like potential in two dimensions. On the other hand, the effective potential (\ref{veff}) also can identified with a Kratzer-Fues-like potential. This potential is a particular case of the Mie potential, widely used to describe interactions in diatomic molecules. This analogy underscores that the developed formalism may have relevance beyond LV contexts, contributing to the understanding of systems exhibiting analogous behaviors.

\section{Qualitative study of the effective potential}
\label{quali_study}

The effective potential has various profiles depending on the values of the parameters 
$\gamma_{l}$ and $\delta$. The profiles can be classified as:
\begin{itemize}
\item Class $1$: $\gamma_{l}^{2}>1/4$ and $\delta>0$ - repulsive Coulomb-like potential (scattering states expected).

\item Class $2$: $\gamma_{l}^{2}>1/4$ and $\delta<0$ - Coulomb-like potential with a well structure (scattering and bound states expected).

\item Class $3$: $\gamma_{l}^{2}>1/4$ and $\delta=0$ - similar to Class 1.

\item Class $4$: $\gamma_{l}^{2}<1/4$ and $\delta>0$ - Coulomb barrier-like potential (scattering and bound states expected).

\item Class $5$: $\gamma_{l}^{2}<1/4$ and $\delta<0$ - attractive Coulomb-like potential (scattering and bound states expected).

\item Class $6$: $\gamma_{l}^{2}<1/4$ and $\delta=0$ - similar to Class 5. 

\item Class $7$: $\gamma_{l}^{2}=1/4$ and $\delta>0$ - similar to Class 1.

\item Class $8$: $\gamma_{l}^{2}=1/4$ and $\delta<0$ - similar to Class 5.
\end{itemize}

\begin{figure}[ht]
\centering
\includegraphics[width=0.93\linewidth,angle=0]{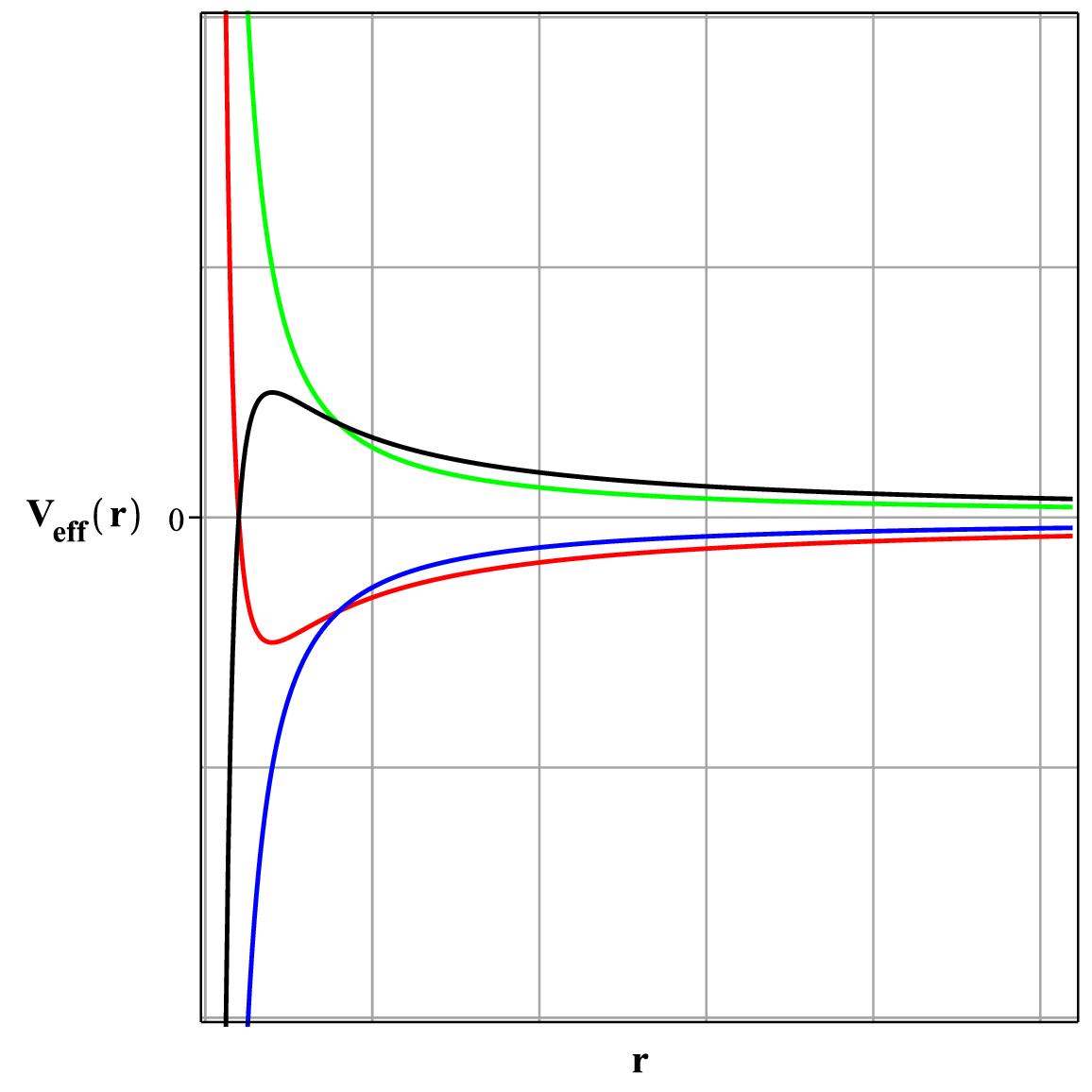}
\caption{Profiles of effective potential: Classes 1, 3 and 7 (green line), class 2 (red line), class 4 (black line), and classes 5, 6 and 8 (blue line).}
\label{f_perfis}
\end{figure}

\begin{figure}[ht]
\centering
\includegraphics[width=0.93\linewidth,angle=0]{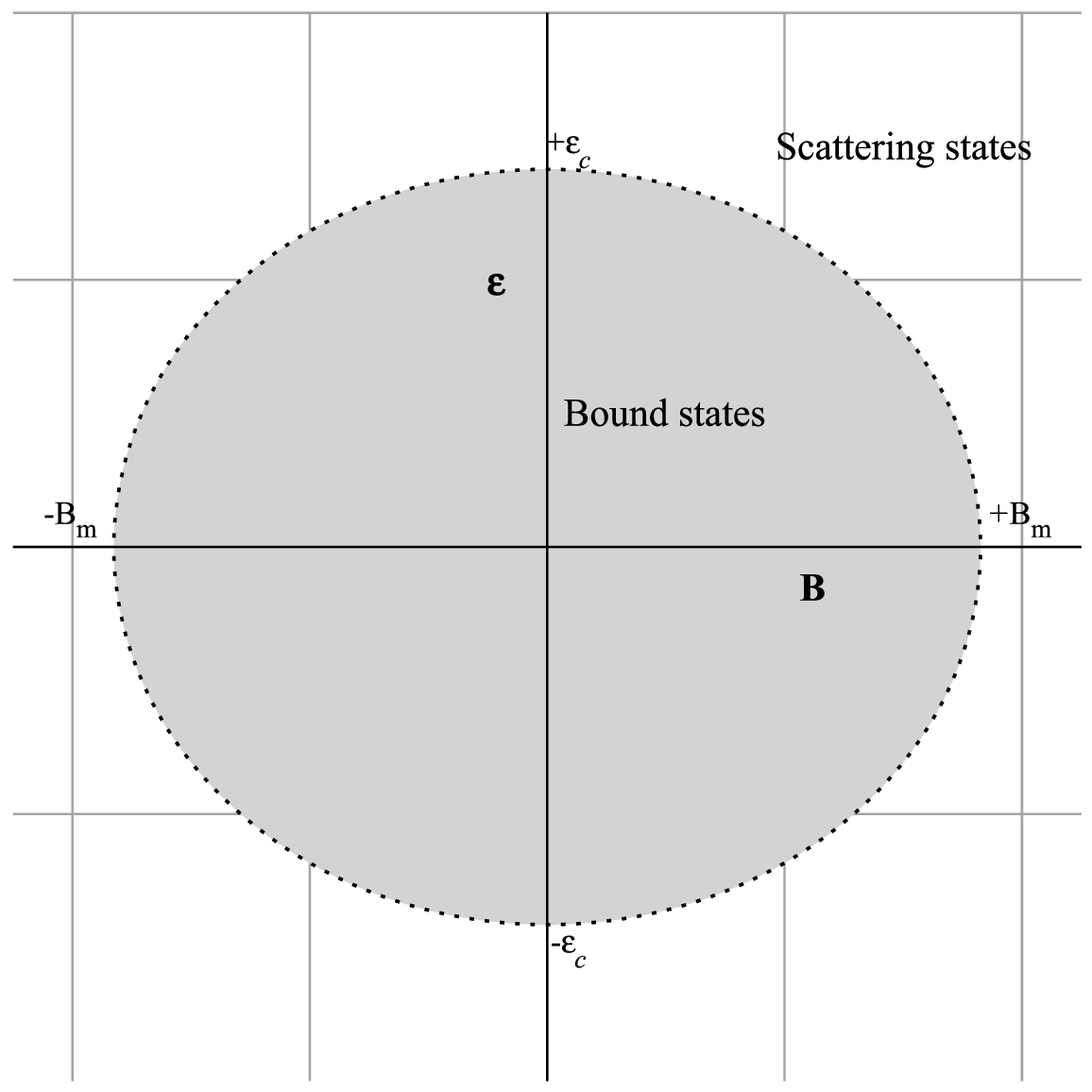}
\caption{Possible regions of bound and scattering states for $\kappa_{2}g>0$.}
\label{figEvsB1}
\end{figure}

\begin{figure}[ht]
\centering
\includegraphics[width=0.93\linewidth,angle=0]{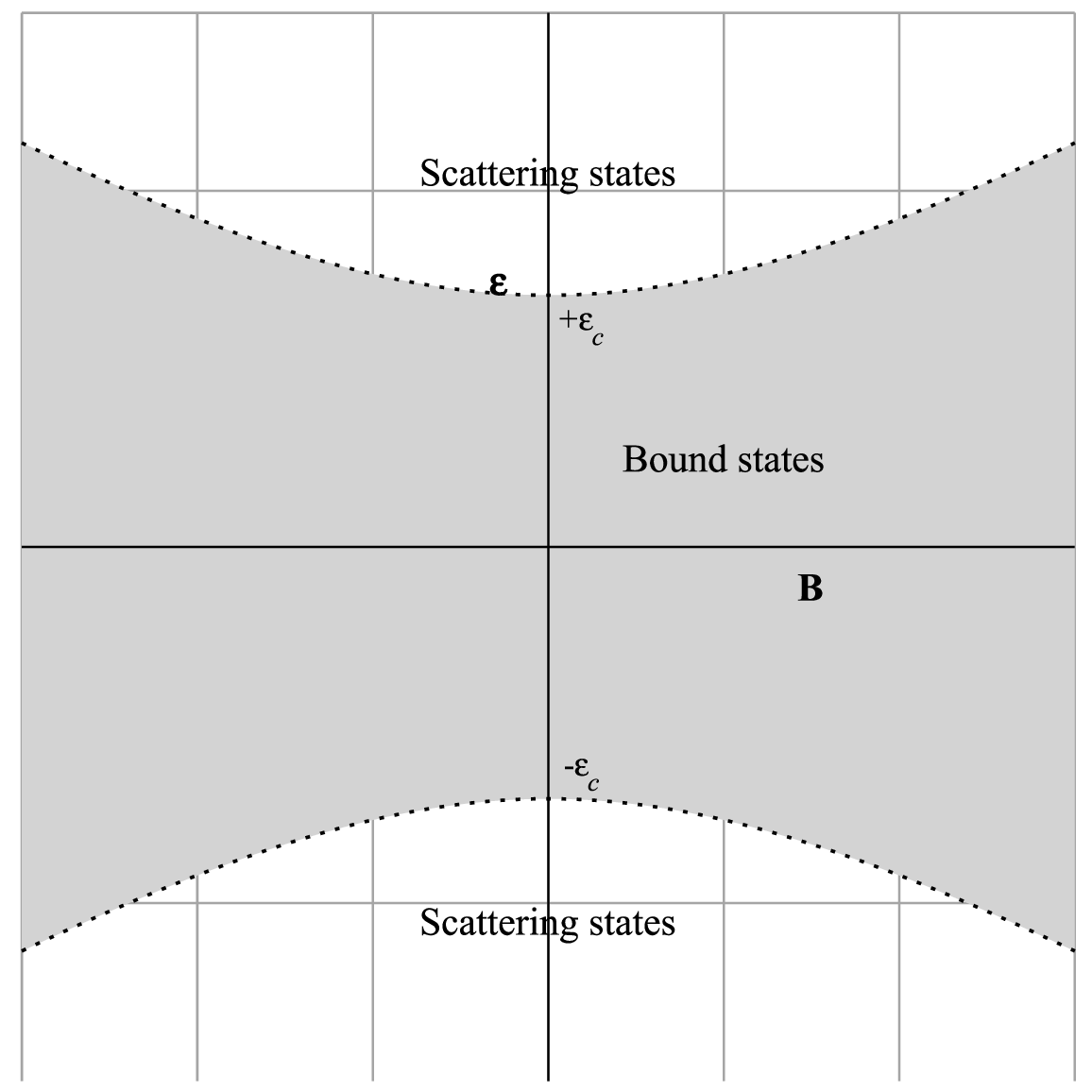}
\caption{Possible regions of bound and scattering states for $\kappa_{2}g<0$.}
\label{figEvsB2}
\end{figure}

\noindent The potential profiles are shown in figure \ref{f_perfis}. We can use this similarity to analyze the scattering and/or bound states for these effective potentials. For all classes, the solution $\phi(r)$ has asymptotic behavior $e^{\pm i\Lambda r}$, so we can see that scattering states only occur if $\Lambda \in \mathbb{R}$, which implies that $\varepsilon> \varepsilon^+$ and $\varepsilon < \varepsilon^-$, where  
\begin{equation}\label{varepsilonpm}
\varepsilon^\pm = \pm \sqrt{m^2+k^2_z-\frac{\kappa_{2}gB^{2}}{2}}\,.
\end{equation} 
\noindent Whereas for classes $2$, $4$, $5$, $6$ and $8$, bound states might occur only if $\Lambda= \pm i |\Lambda|$, i.e. $ \varepsilon^-<\varepsilon< \varepsilon^+$. Depending on the signal of the product $\kappa_{2}g$, we have two possible scenarios:
\begin{itemize}
\item If $\kappa_{2}g>0$, $\varepsilon^{+}$ and $\varepsilon^{+}$ form an ellipse centered at the origin ($\varepsilon=0,B=0$) with width $2\varepsilon_{c}$ and height $2B_{m}$. In this case, figure \ref{figEvsB1} shows the hatched region, which indicates the possible existence of bound states. Conversely, the unhatched region corresponds to condition where scattering states are expected.

\item If $\kappa_{2}g<0$, $\varepsilon^{+}$ and $\varepsilon^{+}$ form an hyperbola centered at the origin ($\varepsilon=0,B=0$) with transverse axis $2\varepsilon_{c}$ and conjugate axis $2B_{m}$. In figure \ref{figEvsB2}, the hatched region represents where bound states are likely to occur and the unhatched regions denote to conditions favourable for scattering states.
\end{itemize}
\noindent Here it was used,
\begin{eqnarray}
\varepsilon_{c} &=& \sqrt{m^{2}+k_{z}^{2}}\,,\label{varepsilonc}\\
B_{m} &=& \sqrt{\frac{2}{\vert\kappa_{2}g\vert}\left(m^{2}+k_{z}^{2}\right)}\,.\label{bmax}
\end{eqnarray}


\noindent Furthermore, it is necessary to analyze additional conditions associated to each effective potential for the possible existence of bound-state solutions. For the class $2$, the effective potential has a well in its structure when $\delta<0$ and $\gamma_{l}^{2}>1/4$, which can favour bound-state solutions. Restringing our study to $B>0$, $g>0$ and $\lambda>0$, the condition $\delta<0$ implies that $\kappa_{3}<0$. It contradicts the results found in Ref.~\cite{EPL136:41002:2021}, where bound-state solutions were obtai\-ned without the requirement of a potential with a well structure. From (\ref{gammal}), the condition $\gamma_{l}^{2}>1/4$ is satisfied for
\begin{equation}\label{cond_gamma1}
\kappa_{1}g>-\frac{2}{\lambda^{2}}\left( l^{2}-\frac{1}{4} \right)\,.
\end{equation}
\noindent On the other hand, if $\kappa_{1}=0$ one obtains $\gamma_{l}^{2}=l^{2}$, which implies that $\vert l\vert>1/2$, i.e. $\vert l\vert \geqslant 1$. 

A different result is obtained for the class $4$, where the effective potential is a Coulomb barrier-like potential when $\delta>0$ ($\kappa_{3}>0$) and $\gamma_{l}^{2}<1/4$. For the class $5$, where the effective potential is a attractive Coulomb-like potential when $\delta<0$ ($\kappa_{3}<0$) and $\gamma_{l}^{2}<1/4$. For the class $8$, the effective potential is similar to class $5$, but in this case $\delta<0$ ($\kappa_{3}<0$) and $\gamma_{l}^{2}=1/4$. These results highlight the crucial role that the sign of $\kappa_{3}$ plays in the possible existence of bound-state solutions for the classes $2$, $4$, $5$ and $8$. From (\ref{delta}), it is evident that the presence of magnetic and electric fields are also crucial to obtain bound-state solutions. Note that the condition $\gamma_{l}^{2}=1/4$ for the class $8$ implies that
\begin{equation}\label{cond_gamma3}
\kappa_{1}g=-\frac{2}{\lambda^{2}}\left( l^{2}-\frac{1}{4} \right)\,.
\end{equation}
\noindent  Finally, for the class $6$ the effective potential is a attractive Coulomb-like potential when $\delta=0$ ($\kappa_{3}=0$) and $\gamma_{l}^{2}<1/4$. Note that the condition $\gamma_{l}^{2}<1/4$ is a common condition for the cases $4$, $5$ and $6$ and it implies that
\begin{equation}\label{cond_gamma2}
\kappa_{1}g<-\frac{2}{\lambda^{2}}\left( l^{2}-\frac{1}{4} \right)\,.
\end{equation} 
\noindent For the special case $\kappa_{1}=0$ one obtains that the condition $\gamma_{l}^{2}<1/4$ is only satisfied when $l=0$. It is worthwhile to mention that the singularity at $r=0$ for $\gamma_{l}^{2}<1/4$ of the cases $4$, $5$ and $6$ menaces the particle to collapse to the center \cite{LANDAU1958} so that the condition $1/4-\gamma_{l}^{2}\leqslant\alpha_{\mathrm{crit}}=1/4$, i.e $\gamma_{l}^{2}\geqslant 0$ ($\kappa_{1}\geqslant 0$) is required for the formation of bound-state solutions. From this qualitative study, it can be concluded that the presence of $\kappa_{2}$ is not essential to study the different effective potential profiles, but it is important to define the allowed interval of the energy for scattering and bound states.

Additionally, when $\gamma_{l}$ becomes imaginary the possibility of spontaneous production of pairs of particles induced by a Coulomb-like field \cite{TMP116:956:1998,TMP158:210:2009,IJTP62:136:2023} is expected. In this case, one obtains that $\kappa_{1}$ needs to be negative and $\vert\kappa_{1}\vert g>2l^{2}/\lambda^{2}$. This problem is addressed in Ref. \cite{EPL145:40002:2024}. It is worth mentioning, that if we consider the $\kappa_{i}$ parameters to be very small, the term $B_{m}$ approaches infinity. In this scenario, the bound states are restricted within the energy gap defined by $+\varepsilon_{c}$ and $-\varepsilon_{c}$, whereas the scattering states are found outside this gap. In other words, our system would converge to the behavior of a relativistic free particle, making it unnecessary to impose constraints on $\kappa_1 g$.

Now let us consider a quantitative treatment of our problem, by examining both scattering and bound states.

\section{Scattering states for $\delta\neq 0$}
\label{kg_sc}

Out of all the cases mentioned in the qualitative study, we will direct our efforts towards class 2. Using the abbreviation
\begin{equation}
\eta=\frac{\delta}{2\Lambda} ,
\end{equation} 
and the change $\rho=-2i\Lambda r$, the equation (\ref{eq4}) becomes
 \begin{equation}
    \frac{d\phi(\rho)}{d\rho^2}+\left(-\frac{1}{4}-\frac{i\eta}{\rho}+\frac{\frac{1}{4}-\gamma_l^2 }{\rho^2} \right)\phi(\rho) =0 . \label{eq5}
\end{equation}
\noindent This second-order differential equation is the called Whittaker
equation, which admits two linearly independent solutions  $M_{-i\eta,\gamma_l}(\rho)$ and $W_{-i\eta,\gamma_l}(\rho)$ behaving like $\rho^{1/2+\gamma_l}$ and $\rho^{1/2-\gamma_l}$ close to the origin, respectively. Owing to the condition $\phi(0)=0$, only the regular solution at the origin is allowed. In this case, the solution is given by
\begin{equation}
\phi(\rho)= C_1 \, \rho^{\gamma_l+1/2} e^{-\rho/2} M\left(\gamma_l+i\eta+\frac{1}{2},1+2\gamma_l;\rho\right)\,, \label{phi1}
\end{equation}
\noindent where $C_1$ is a arbitrary constant and $M(a,b; \rho)$ is the Kummer's function, whose asymptotic behavior for large $|\rho|$ with a purely imaginary $\rho=-i|\bar{\rho}|$, where $|\bar{\rho}|=2\Lambda r$ with $\Lambda>0$ is given by \cite{JAMES1991} 
\begin{equation}\label{com_as}
\begin{split}
M(a,b;\rho) & \simeq \frac{\Gamma(b)}{\Gamma(a-b)}e^{-\frac{i\pi a}{2}}|\bar{\rho}|^{-a}\\
&+\frac{\Gamma(b)}{\Gamma(a)}e^{-i[|\bar{\rho}|-\frac{\pi(b-a)}{2}]}|\bar{\rho}|^{a-b}.
\end{split}
\end{equation}
\noindent Using this last result, we can show that for $\vert\rho\vert\gg 1$, the asymptotic behavior of (\ref{phi1}) is given by \cite{AP146:1:1983}
\begin{equation}
\phi(r)\simeq \cos \left(\Lambda r-\frac{\vert l\vert\pi}{2}-\frac{\pi}{4}+\delta_l\right),
\end{equation}
\noindent where the relativistic phase shift $\delta_l = \delta_l(\eta)$ is given by
\begin{equation}\label{phase_shift}
\delta_l=\frac{\pi}{2}(\vert l\vert-\gamma_l)+ \text{arg}\left[\Gamma \left(\frac{1}{2}+\gamma_l+i\eta\right)\right]\,.
\end{equation}
\noindent In time-independent scattering theory, the solution of the KG equation (\ref{eq3}) has the asymptotic form \cite{AP146:1:1983}
\begin{equation}
\psi(\vec{r})\simeq \mathrm{e}^{i\Lambda r\cos\varphi}+f(\Lambda,\varphi)\frac{\mathrm{e}^{i\Lambda r}}{\sqrt{r}}\,,
\end{equation}
\noindent where the first term represents a plane wave moving along the direction $\varphi=0$ toward the scatterer, and the second term represents a radially outgoing wave. We can write the scattering amplitude as a partial wave series
\begin{equation}
f(\Lambda,\varphi)=\sum_{l=-\infty}^{\infty}f_{l}(\Lambda)\mathrm{e}^{il\varphi}\,,
\end{equation}
\noindent where the partial scattering amplitude is
\begin{equation}
f_{l}(\Lambda)=\frac{\mathrm{e}^{2i\delta_{l}}-1}{\sqrt{2\pi i\Lambda}}\,,
\end{equation}
\noindent with the phase shift (\ref{phase_shift}), up to a logarithmic phase inherent to the Coulomb-like field. From this last expression, one can recognize the scattering $S$-matrix as
\begin{equation}\label{s_matrix}
S_{l}=\mathrm{e}^{2i\delta_{l}}=\mathrm{e}^{i\pi(\vert l\vert-\gamma_{l})}\frac{\Gamma\left(1/2+\gamma_{l}+i\eta\right)}{\Gamma\left(1/2+\gamma_{l}-i\eta\right)}\,.
\end{equation} 
\noindent 

\section{Bound states for $\delta\neq 0$}
\label{kg_bs}

The energies of the bound-state solutions can be obtained from poles of the $S$-matrix when one considers $\Lambda$ imaginary. Then, if $\Lambda=i\vert \Lambda\vert$, the $S$-matrix becomes infinite when $1/2+\gamma_{l}+i\eta=-n$, where $n=0,1,2,\ldots$, owing to the poles of the gamma function in the numerator of (\ref{s_matrix}). From the qualitative study, bound-state solutions are possible only for $\kappa_{3}<0$, corresponding to energies in the interval $\varepsilon^{-}<\varepsilon<\varepsilon^{+}$ and the spectrum is given by
\begin{equation}\label{ener1}
\varepsilon=\pm\sqrt{m^{2}+k_{z}^{2}-\frac{\kappa_{2}gB^{2}}{2}-\frac{(\kappa_{3}g)^{2}\lambda^{2}B^{2}}{4\mu^{2}}}\,,
\end{equation}  
\noindent with $\mu=n+1/2+\gamma_{l}$. Note that the energy spectrum in equation (\ref{ener1}) aligns with Eq.~(13) of Ref.~\cite{EPL136:41002:2021}. It can be observed that the energy spectrum is symmetrical about $\varepsilon=0$, encompassing both particle energies (positive energy) and antiparticle energies (negative energy). From (\ref{ener1}), one can see that the energy spectrum is irrespective to the sign of $\kappa_{3}$. Furthermore, note that the accidental degeneracy exists when $\kappa_{1}=0$ ($\gamma_{l}=\vert l\vert$) and in this case the energy $\varepsilon$ depends on $n$ and $\vert l\vert$ via the combination $n+\vert l\vert$, in such a way that each energy is $2\mu$-fold degenerate, where in this case we have that $\mu=n+1/2+\vert l\vert$. 

\begin{figure}[ht]
\centering
\includegraphics[width=0.93\linewidth,angle=0]{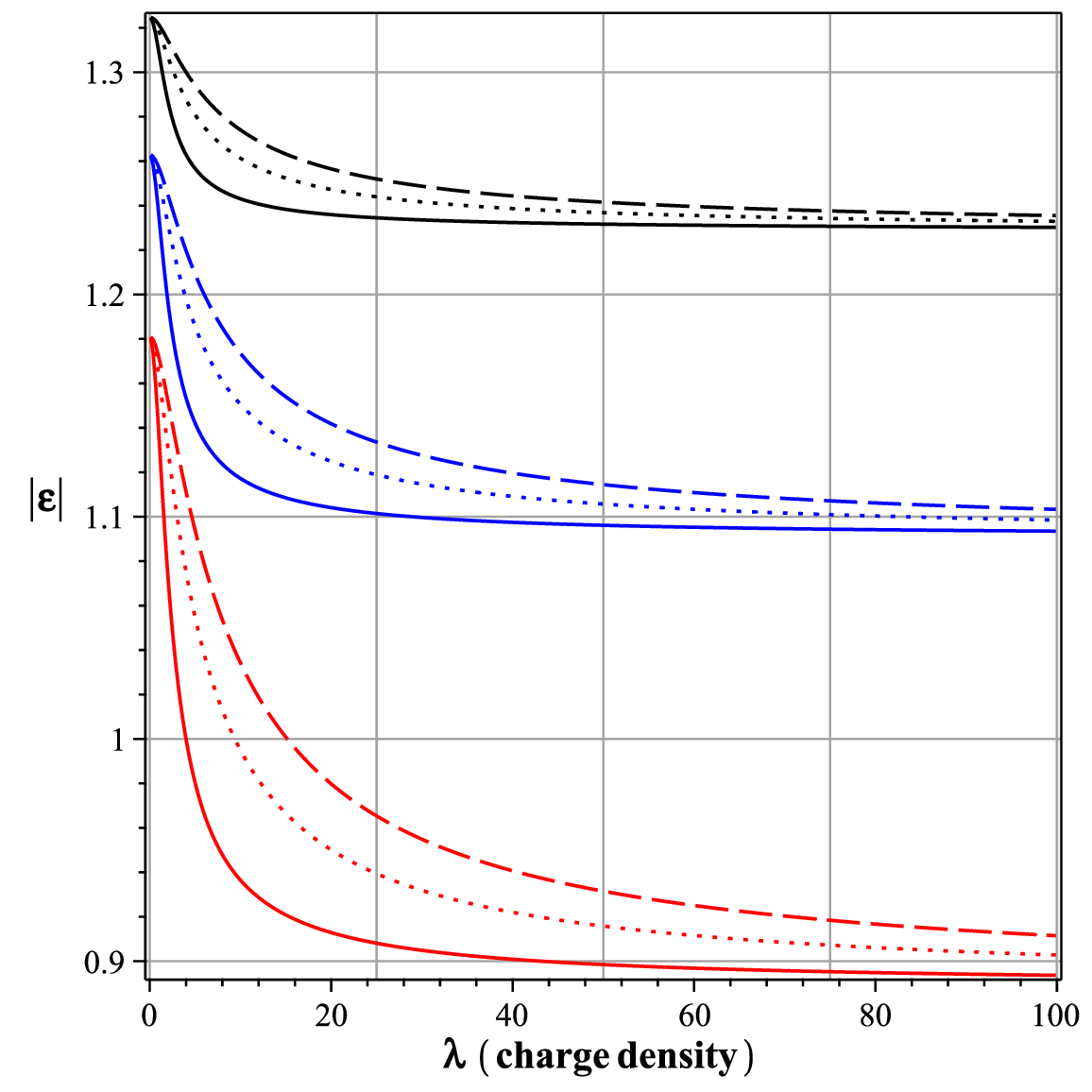}
\caption{Plots of the energy as function of $\lambda$ for $|l|=1$ and different values of $n$ and $B$. For $n=0$ [solid line], $n=1$ [dotted line] and $n=2$ [dashed line]. For $B=0.7$ [black], $B=0.9$ [blue] and $B=1.1$ [red].}
\label{EvsLambda1}
\end{figure}

Figure \ref{EvsLambda1} depicts the profile of the absolute energy values as a function of charge density $\lambda$ for $\vert l\vert=1$. In this figure, we choose $m=k_{z}=g=1$, $\kappa_{1}=\kappa_{2}=1$ and $\kappa_{3}=-1$, and examine the first three quantum numbers and three different values of the magnetic field $B$. It is evident that for fixed values of $n$ and $\vert l\vert$, the energy $\vert\varepsilon\vert$ decreases as $B$ increases, in agreement with Fig. \ref{EvsLambda1}. Additionally, for a fixed charge density, the spacing between energy levels increases as $B$ increases. From (\ref{ener1}), can be identified a critical value for the magnetic field $B_{c}<B_{m}$, which makes the energy spectrum complex. In this case, the critical magnetic field is given by
\begin{equation}\label{cmc}
B_{c}=\sqrt{\frac{m^{2}+k_{z}^{2}}{\frac{\kappa_{2}g}{2}+\frac{(\kappa_{3}g)^{2}\lambda^{2}}{4\mu}}}\,,
\end{equation}   
\noindent which dependent of $n$ and $l$. This result shows us that the magnetic field cannot be too intense ($B<B_{c}$). Furthermore, both figures show that as the charge density increases, the energy levels for a fixed magnetic field converge to a constant value. This constant value can be obtained by applying the limit $\lambda\gg 1$ to equation (\ref{ener1}). In this limit, the energy spectrum simplifies to
\begin{equation}\label{ener_l}
\vert\varepsilon\vert\simeq\sqrt{m^{2}+k_{z}^{2}-\frac{\kappa_{2}gB^{2}}{2}-\frac{(\kappa_{3}g)^{2}B^{2}}{2\kappa_{1}g}}\,,
\end{equation}
\noindent which is independent of $n$ and $l$, and depends solely on the magnetic field. Lastly, if the quantum number $\vert l\vert$ increases, the energy curves become steeper, which indicates a sharper change in energy levels.

The solution for all $r$ can be written as
\begin{equation}\label{solu}
\phi(\rho)=N_{n} r^{\gamma_{l}+1/2}\mathrm{e}^{-\vert\Lambda\vert r}L^{(2\gamma_{l})}_{n}(2\vert \Lambda\vert r)\,,
\end{equation}
\noindent due to $M(-n,b;\rho)$ with $b>0$ is proportional to the generalized Laguerre polynomial $L^{(b-1)}_{n}(\rho)$ \cite{ABRAMOWITZ1965}, a polynomial of degree $n$ with $n$ distinct positive zeros in the range $[0,\infty)$. The normalization constant $N_{n}$ in (\ref{solu}) is obtained by means of the normalization condition. The charge density $J^{0}$ (\ref{j0}) dictates that $\phi$ must be normalized as
\begin{equation}\label{norma_c}
\frac{\vert\varepsilon\vert}{m}\int dr \vert\phi\vert^{2}=1\,.
\end{equation}
\noindent In this way, the normalization constant can be written as
\begin{equation}\label{cn1}
N_{n}=\sqrt{\frac{m(2\vert\Lambda\vert)^{2\gamma_{l}+2}n!}{\vert\varepsilon\vert(2n+2\gamma_{l}+1)[(n+2\gamma_{l})!]^{3}}}\,,
\end{equation} 
\noindent with $\vert\varepsilon\vert\neq 0$.

\section{Conclusions}
\label{kg_c}

We have analyzed the effects of LSV on relativistic neutral scalar bosons via KG formalism, focusing on both scattering and bound states. In this letter, we have been considered static electromagnetic fields caused by LSV in the parity-even sector of the CPT-even photon sector in the SME. By choosing a cross-configuration of an inhomogeneous static electric field and homogeneous static magnetic field together with non-null components of the tensor $(K_F)_{\mu \nu \alpha \beta}$ given by $\kappa_{1}$, $\kappa_{2}$ and $\kappa_{3}$, the radial part of the modified Klein-Gordon equation for neutral scalar bosons in cylindrical coordinates can be expressed as a time-independent radial Schrödinger equation for a Coulomb-like potential in two dimensions. By performing a detailed qualitative analysis of the effective potential, it has been found that the effective potential has eight profiles (eight classes) depending on the values of the parameters $\gamma_{l}$ and $\delta$ defined by equations (\ref{delta}) and (\ref{gammal}), respectively. These eight profiles may support bound states and/or scattering states depending on whether the constrains between the potential parameters are satisfied. From this qualitative analysis, we conclude that while the presence of $\kappa_{2}$ is not critical for examining the various profiles of the effective potential, it plays a key role in determining the allowed interval of the energy for both scattering and bound states. This qualitative analysis is crucial as it provides a clear understanding of the possible energy ranges associated with bound and scattering states. Moreover, it highlights the broader complexity and richness of the problem.

Restricting our study to class $2$ ($\gamma_{l}^{2}>1/4$ and $\delta<0$), scattering solutions can be studied by solving a Whittaker differential equation. The relativistic phase shift, scattering amplitude and scattering $S$-matrix have been calculated as a function of the potential parameters. In this case, scattering states only occur if $\Lambda \in \mathbb{R}$, which implies that $\varepsilon> \varepsilon^+$ and $\varepsilon < \varepsilon^-$, where $\varepsilon^{\pm}$ is given by expression (\ref{varepsilonpm}). The poles of $S$-matrix when $\Lambda=i\vert\Lambda\vert$ provided bound-state solutions only for $\delta=\kappa_{3}g\lambda B<0$. Restringing our study to $B>0$, $g>0$ and $\lambda>0$, the condition $\delta<0$ implies that $\kappa_{3}<0$. From this result, we have concluded that the presence of $B$, $g$, $\lambda$ and $\kappa_{3}$ is crucial for the existence of bound-state solutions. Although $\kappa_{1}$ does not play a fundamental role in the existence of bound states, it is highly significant in the context of pair production, as recently discussed in reference \cite{EPL145:40002:2024}.

We have demonstrated that the energy spectrum lies within the interval $\varepsilon^{-} < \varepsilon < \varepsilon^{+}$, where both particle and antiparticle energy levels are part of the spectrum. It is important to highlight that while the energy expression aligns with the result obtained in reference \cite{EPL136:41002:2021}, no parameter restrictions were established in that reference. Notably, the particle and antiparticle spectra are symmetric with respect to $\varepsilon = 0$, indicating the absence of any channel for spontaneous boson-antiboson pair creation. We have also identified a critical value for the magnetic field, $B_{c}$, beyond which the energy spectrum becomes complex. This result indicates that the magnetic field cannot be too intense. On the other hand, the charge density $\lambda$ does not impose any restriction on its value. In fact, for $\lambda\gg 1$ with values of $g$, $\kappa_{1}$, $\kappa_{2}$ e $\kappa_{3}$ fixed, the energy spectrum approaches a constant value that depends solely on the magnetic field $B$, becoming independent of the quantum numbers $n$ and $l$. Finally, we derived the normalized wave function using the correct procedure, which differs from the approach presented in reference \cite{EPL136:41002:2021}. 

Our work goes beyond merely extending the results of reference \cite{EPL136:41002:2021} by including the study of scattering states. We also conduct an in-depth analysis of bound states, providing significant new insights, challenging previous findings and uncovering previously unexplored results not addressed in reference \cite{EPL136:41002:2021}. 

Additionally, the effective potential considered in this work (Class 2), as illustrated in Figure \ref{f_perfis}, exhibits a stability point characterized by a well-like structure, as discussed in the text. From a classical perspective, small oscillations around this point can be approximated as harmonic. From a quantum mechanical perspective, the presence of bound states will depend on the size and depth of the effective potential near this point. While this intriguing question is beyond the scope of the present work, it undoubtedly offers an interesting direction for future research and certainly merits further investigation.

This work contributes to the understanding of LSV effects in relativistic quantum systems, with potential implications for fundamental physics.

\acknowledgments
The authors express their gratitude to the anonymous reviewers for their valuable comments, which significantly enhanced the quality of this paper. L. B. Castro also extends his thanks to R. Casana for the insightful discussions. This work was supported in part by means of funds provided by CNPq, Brazil, Grant No. 308172/2023-0 (PQ), CAPES - Finance code 001, Brazil and FAPEMA, Brazil.

\bibliographystyle{eplbib}


\begin{thebibliography}{10}
\expandafter\ifx\csname url\endcsname\relax\def\url#1{\texttt{#1}}\fi

\bibitem{PRL63:224:1989}
\Name{Kosteleck\'y V.~A. \and Samuel S.} \REVIEW{Phys. Rev.
  Lett.}{63}{1989}{224}.

\bibitem{PRD39:683:1989}
\Name{Kosteleck\'y V.~A. \and Samuel S.} \REVIEW{Phys. Rev. D}{39}{1989}{683}.

\bibitem{PRD40:1886:1989}
\Name{Kosteleck\'y V.~A. \and Samuel S.} \REVIEW{Phys. Rev. D}{40}{1989}{1886}.

\bibitem{PRL66:1811:1991}
\Name{Kosteleck\'y V.~A. \and Samuel S.} \REVIEW{Phys. Rev.
  Lett.}{66}{1991}{1811}.

\bibitem{NPB359:545:1991}
\Name{{Alan Kosteleck\'y} V. \and Potting R.} \REVIEW{Nucl. Phys.
  B}{359}{1991}{545}.

\bibitem{PRD51:3923:1995}
\Name{Kosteleck\'y V.~A. \and Potting R.} \REVIEW{Phys. Rev.
  D}{51}{1995}{3923}.

\bibitem{PLB381:89:1996}
\Name{{Alan Kosteleck\'y} V. \and Potting R.} \REVIEW{Phys. Let.
  B}{381}{1996}{89}.

\bibitem{PRD55:6760:1997}
\Name{Colladay D. \and Kosteleck\'y V.~A.} \REVIEW{Phys. Rev.
  D}{55}{1997}{6760}.

\bibitem{PRD58:116002:1998}
\Name{Colladay D. \and Kosteleck\'y V.~A.} \REVIEW{Phys. Rev.
  D}{58}{1998}{116002}.

\bibitem{PRD59:116008:1999}
\Name{Coleman S. \and Glashow S.~L.} \REVIEW{Phys. Rev. D}{59}{1999}{116008}.

\bibitem{PRD41:1231:1990}
\Name{Carroll S.~M., Field G.~B. \and Jackiw R.} \REVIEW{Phys. Rev.
  D}{41}{1990}{1231}.

\bibitem{PRL87:251304:2001}
\Name{Kosteleck\'y V.~A. \and Mewes M.} \REVIEW{Phys. Rev.
  Lett.}{87}{2001}{251304}.

\bibitem{PRD66:056005:2002}
\Name{Kosteleck\'y V.~A. \and Mewes M.} \REVIEW{Phys. Rev.
  D}{66}{2002}{056005}.

\bibitem{PRL97:140401:2006}
\Name{Kosteleck\'y V.~A. \and Mewes M.} \REVIEW{Phys. Rev.
  Lett.}{97}{2006}{140401}.

\bibitem{AP360:596:2015}
\Name{Bakke K. \and Belich H.} \REVIEW{Ann. Phys. (N.Y.)}{360}{2015}{596}.

\bibitem{AP373:115:2016}
\Name{Bakke K. \and Belich H.} \REVIEW{Ann. Phys. (N.Y.)}{373}{2016}{115}.

\bibitem{EPJP132:25:2017}
\Name{Vit\'{o}ria R. L.~L., Belich H. \and Bakke K.} \REVIEW{Eur. Phys. J.
  Plus}{132}{2017}{25}.

\bibitem{EPJC78:999:2018}
\Name{Vit\'{o}ria R. L.~L. \and Belich H.} \REVIEW{Eur. Phys. J.
  C}{78}{2018}{999}.

\bibitem{AP399:117:2018}
\Name{Vit\'oria R., Bakke K. \and Belich H.} \REVIEW{Ann. Phys.
  (N.Y.)}{399}{2018}{117}.

\bibitem{ADHEP2019:8462973:2019}
\Name{Vit\'oria R. L.~L. \and Belich H.} \REVIEW{AdHEP}{2019}{2019}{8462973}.

\bibitem{ADHEP2019:1248393:2019}
\Name{Vit\'oria R. L.~L. \and Belich H.} \REVIEW{AdHEP}{2019}{2019}{1248393}.

\bibitem{IJMPA35:2050023:2020}
\Name{Bakke K. \and Belich H.} \REVIEW{Int. J. Mod. Phys.
  A}{35}{2020}{2050023}.

\bibitem{EPJP135:123:2020}
\Name{Vit\'{o}ria R. L.~L. \and Belich H.} \REVIEW{Eur. Phys. J.
  Plus}{135}{2020}{123}.

\bibitem{EPL136:41002:2021}
\Name{Ahmed F.} \REVIEW{EPL}{136}{2021}{41002}.

\bibitem{EPL136:61001:2021}
\Name{Ahmed F.} \REVIEW{EPL}{136}{2022}{61001}.

\bibitem{EPL139:30001:2022}
\Name{Ahmed F.} \REVIEW{EPL}{139}{2022}{30001}.

\bibitem{EPL141:60002:2023}
\Name{Vicente A. G.~J., Castro L.~B. \and Obispo A.~E.}
  \REVIEW{EPL}{141}{2023}{60002}.

\bibitem{PRD81:105015:2010}
\Name{Casana R., Ferreira M.~M. \and Silva M. R.~O.} \REVIEW{Phys. Rev.
  D}{81}{2010}{105015}.

\bibitem{PRD84:045008:2011}
\Name{Casana R., Carvalho E.~S. \and Ferreira M.~M.} \REVIEW{Phys. Rev.
  D}{84}{2011}{045008}.

\bibitem{JPG39:055004:2012}
\Name{Bakke K., Silva E.~O. \and Belich H.} \REVIEW{J. Phys. G: Nucl. Part.
  Phys.}{39}{2012}{055004}.

\bibitem{EPJP135:247:2020}
\Name{Vit\'{o}ria R. L.~L. \and Belich H.} \REVIEW{Eur. Phys. J.
  Plus}{135}{2020}{247}.

\bibitem{LANDAU1958}
\Name{Landau L.~D. \and Lifshitz E.~M.} \Book{Quantum Mechanics} (Pergamon, New
  York) 1958.

\bibitem{TMP116:956:1998}
\Name{Khalilov V.~R.} \REVIEW{Theor. Math. Phys.}{116}{1998}{956}.

\bibitem{TMP158:210:2009}
\Name{Khalilov V.~R.} \REVIEW{Theor. Math. Phys.}{158}{2009}{210}.

\bibitem{IJTP62:136:2023}
\Name{Belbaki B. \and Bounames A.} \REVIEW{Int. J. Theor.
  Phys.}{62}{2023}{136}.

\bibitem{EPL145:40002:2024}
\Name{Jir\'{o}n A.~G., Obispo A.~E., Loayza J. D.~E., Quispe J.~C. \and Castro
  L.~B.} \REVIEW{EPL}{145}{2024}{40002}.

\bibitem{JAMES1991}
\Name{Seaborn J.~B.} \Book{Hypergeometric Functions and Their Applications}
  (Springer - Verlag, New York) 1991.

\bibitem{AP146:1:1983}
\Name{Ruijsenaars S.} \REVIEW{Ann. Phys. (N.Y.)}{146}{1983}{1}.

\bibitem{ABRAMOWITZ1965}
\Name{Abramowitz M. \and Stegun I.~A.} \Book{Handbook of Mathematical
  Functions} (Dover, Toronto) 1965.

\end{thebibliography}

\end{document}